# CR-MAC: A Multichannel MAC Protocol for Cognitive Radio Ad Hoc Networks

S. M. Kamruzzaman

School of Electronics and Information Engineering
Hankuk University of Foreign Studies, Korea
smzaman@hufs.ac.kr

## ABSTRACT

*This paper proposes a cross-layer based cognitive radio multichannel medium access control (MAC) protocol with TDMA, which integrate the spectrum sensing at physical (PHY) layer and the packet scheduling at MAC layer, for the ad hoc wireless networks. The IEEE 802.11 standard allows for the use of multiple channels available at the PHY layer, but its MAC protocol is designed only for a single channel. A single channel MAC protocol does not work well in a multichannel environment, because of the multichannel hidden terminal problem. Our proposed protocol enables secondary users (SUs) to utilize multiple channels by switching channels dynamically, thus increasing network throughput. In our proposed protocol, each SU is equipped with only one spectrum agile transceiver, but solves the multichannel hidden terminal problem using temporal synchronization. The proposed cognitive radio MAC (CR-MAC) protocol allows SUs to identify and use the unused frequency spectrum in a way that constrains the level of interference to the primary users (PUs). Our scheme improves network throughput significantly, especially when the network is highly congested. The simulation results show that our proposed CR-MAC protocol successfully exploits multiple channels and significantly improves network performance by using the licensed spectrum band opportunistically and protects PUs from interference, even in hidden terminal situations.*

## KEYWORDS

*Cognitive radio, multichannel MAC, ad hoc networks, frequency spectrum, TDMA, channel sensing.*

## 1. INTRODUCTION

Cognitive radio (CR) has emerged as the solution to the problem of spectrum scarcity for wireless applications. It has been using the vacant spectrum of licensed band opportunistically. Cognitive radio networks (CRNs) refer to networks where nodes are equipped with a spectrum agile radio which has the capabilities of sensing the available spectrum band, reconfiguring radio frequency, switching to the selected frequency band and use it efficiently without interference to PUs [1] [2]. CR Ad hoc networks (CRANs) are emerging, infrastructure less multi-hop CRNs. The CR users (nodes) can communicate with each other through ad hoc connection.

The throughput of multi-hop wireless networks can be significantly improved by multichannel communications compared with single channel since the use of multiple channel can reduce the interference influence [3] [4]. We consider a multichannel CRN, in which every node is equipped with single network interface card (NIC) and can be tuned to one of the available channels. A pair of NICs can communicate with each other if they are on the same channel and are within the transmission range of each other.

Although the basic idea of CR is simple, the efficient design of CRNs imposes the new challenges that are not present in the traditional wireless networks [5]–[7]. Specifically, identifying the time-varying channel availability imposes a number of nontrivial design problems to the MAC layer. One of the most difficult, but important, design problems is how the SUs decide when and which channel they should tune in to in order to transmit/receive the SUs' packets without interference to the PUs. This problem becomes even more challenging in wireless ad hoc networks where there are no centralized controllers, such as base stations or access points.

As CRNs need to use several channels in parallel to fully utilize the spectrum opportunities, the MAC layer should accordingly be designed. Multichannel MAC protocols have clear advantages over single channel MAC protocols: They offer reduced interference among users, increased network throughput due to simultaneous transmissions on different channels, and a reduction of the number of CRs affected by the return of a licensed user [8]. By exploiting multiple channels, we can achieve a higher network throughput than using single channel, because multiple transmissions can take place without interfering. Designing a MAC protocol that exploits multiple channels is not an easy task, due to the fact that each of current IEEE 802.11 device is equipped with one half-duplex transceiver. The transceiver is capable of switching channels dynamically, but it can only transmit or listen on one channel at a time. Thus, when a node is listening on a particular channel, it cannot hear communication taking place on a different channel. Due to this, a new type of hidden terminal problem occurs in this multichannel environment, which is referred as multichannel hidden terminal problem. So, a single channel MAC protocol (such as IEEE 802.11 DCF) does not work well in a multichannel environment where nodes may dynamically switch channels.

To amend the aforementioned problems of the existing schemes, in this paper, we propose a multichannel CR-MAC protocol which enables nodes to dynamically negotiate channels such that multiple communications can take place in the same region simultaneously, each in different channel. The network we consider is an ad hoc network that does not rely on infrastructure, so there is no central authority to perform channel management. The main idea is to divide time in to fixed-time frame intervals, and have a small window at the start of each interval to indicate traffic and negotiate channels and time slots for use during the interval. A similar approach is used in IEEE 802.11 power saving mechanism (PSM) [9], explained in section 3.2. The proposed scheme can eliminates contention between nodes, decomposes contending traffics over different channels and timeslots based on actual traffic demand. As a result, the proposed scheme leads to significant increases in network throughput and decreases the end-to-end delay.

## 2. RELATED WORK

Recently, several attempts were made to develop MAC protocols for CRNs [10]-[18]. One of the key challenges to enabling CR communications is how to perform opportunistic medium access control (MAC) while limiting the interference imposed on PUs. The IEEE 802.22 working group is in the process of standardizing a centralized MAC protocol that enables spectrum reuse by CR users operating on the TV broadcast bands [19]. In [14]-[16], centralized protocols were proposed for coordinating spectrum access. For a CR ad hoc network without centralized control, it is desirable to have a distributed MAC protocol that allows every CR user to individually access the spectrum.

A number of multichannel contention-based MAC protocols were previously proposed in the context of CRNs [10]-[13]. The CRN MAC protocol in [10] jointly optimizes the multichannel power/rate assignment, assuming a given power mask on CR transmissions. How to determine an appropriate power mask remains an open issue. Distance and traffic-aware channel assignment (DDMAC) in cognitive radio networks is proposed in [11]. It is a spectrum sharing protocol for CRNs that attempts to maximize the CRN throughput through a novel probabilistic channel assignment algorithm that exploits the dependence between the signal's attenuation model and the transmission distance while considering the prevailing traffic and interference conditions. A bandwidth sharing approach to improve licensed spectrum utilization (AS-MAC) is presented in [12] is a spectrum sharing protocol for CRNs that coexists with a GSM network. CR users select channels based on the CRN's control exchanges and GSM broadcast information. Explicit coordination with the PUs is required. In [18], the authors developed a spectrum-aware MAC protocol for CRNs (CMAC). CMAC enables opportunistic access and sharing of the available white spaces in the TV spectrum by adaptively allocating the spectrum among contending users.

A cognitive MAC protocol for multichannel wireless networks (C-MAC) is proposed in [20], which operates over multiple channels, and hence is able to effectively deal with the dynamics of resource availability due to PUs and mitigate the effects of distributed quiet periods utilized for PU signal detection. In C-MAC, each channel is logically divided into recurring superframes which, in turn, include a slotted beaconing period (BP) where nodes exchange information and negotiate channel usage. Each node transmits a beacon in a designated beacon slot during the BP, which helps in dealing with hidden nodes, medium reservations, and mobility.

CR based multichannel MAC protocols for wireless ad hoc networks (CRM-MAC) is proposed in [21], which integrate the spectrum sensing and packet scheduling. In their protocols each SU is equipped with two transceivers. One of the transceivers operates on a dedicated control channel, while the other is used as a CR that can periodically sense and dynamically utilize the identified unused channels. CR-enabled multichannel MAC (CREAM-MAC) protocol is proposed in [22], which integrates the spectrum sensing at physical layer and packet scheduling at MAC layer, over the wireless networks. In the proposed CREAM-MAC protocol, each SU is equipped with a CR-enabled transceiver and multiple channel sensors. The proposed CREAM-MAC enables the SUs to best utilize the unused frequency spectrum while avoiding the collisions among SUs and between SUs and PUs.

Distributed CR MAC (COMAC) protocol is presented in [23] that enable unlicensed users to dynamically utilize the spectrum while limiting the interference on PUs. The main novelty of COMAC lies in not assuming a predefined SU-to-PU power mask and not requiring active coordination with PUs. COMAC provides a statistical performance guarantee for PUs by limiting the fraction of the time during which the PUs' reception is negatively affected by CR transmissions. To provide such a guarantee, COMAC develop probabilistic models for the PU-to-PU and the PU-to-SU interference under a Rayleigh fading channel model. From these models, they derive closed-form expressions for the mean and variance of interference.

A distributed multichannel MAC protocol for multi-hop CRNs (MMAC-CR) is proposed in [24] that look at CR-enabled networks with distributed control. In addition to the spectrum scarcity, energy is rapidly becoming one of the major bottlenecks of wireless operations and has to be considered as a key design criterion. They present an energy-efficient distributed multichannel MAC protocol for CR networks. Decentralized cognitive MAC (DC-MAC) for dynamic spectrum access is presented in [25] is a cross-layer distributed scheme for spectrum allocation/sensing. It provides an optimization framework based on partially observable Markov decision processes, with no insights into protocol design, implementation, and performance.

A CR MAC protocol using statistical channel allocation for wireless ad hoc networks (SCA-MAC) is presented in [26]. SCA-MAC is a CSMA/CA based protocol, which exploits statistics of spectrum usage for decision making on channel access. For each transmission, the sender negotiates with the receiver on transmission parameters through the control channel. Synchronized MAC protocol for multi-hop CRNs (SYN-MAC) is proposed in [27], where the use of common control channel (CCC) is avoided. The scheme is applicable in heterogeneous environments where channels have different bandwidths and frequencies of operation.

## 3. BACKGROUND

We first present some background information on the distributed coordination function (DCF) of IEEE 802.11, which is the standard reference for MAC operations in an ad hoc network, and its PSM. At last we discuss the multichannel hidden terminal problem in the end of this section.

### 3.1. IEEE 802.11 Distributed Coordination Function (DCF)

The IEEE 802.11 DCF relies on a continuous sensing of the wireless channel. The algorithm used is called carrier sense multiple access with collision avoidance (CSMA/CA). If a node has a packet to transmit, then it transmits if the medium is sensed to be idle longer than a DCF interframe space (DIFS). If not, then it randomly chooses a backoff value from the interval $[0, W-1]$, where $W$ is defined as the contention window. This backoff counter is decremented every slot after the channel is sensed idle longer than a DIFS. If the backoff counter reaches zero, then the station transmits. Two different intervals, DIFS and SIFS, enable each packet to have different priority when contending for the channel. A node waits for a DIFS before transmitting an RTS, but waits for a SIFS before sending a CTS or an ACK. Thus, an ACK packet will win the channel when contending with RTS or DATA packets because the SIFS duration is smaller than DIFS.

A node is also able to reserve the channel for data transmission by exchanging ready to send (RTS) and clear to send (CTS) packets. If a node has a packet ready for transmission, then it can try to send an RTS frame using the DCF. After receiving an RTS frame, the destination replies with a CTS packet. Both RTS and CTS frames carry the expected duration of transmission. Nodes overhearing this handshake have to defer their transmissions for this duration. For this reason, each host maintains a variable called the network allocation vector (NAV) that records the duration of time it must defer its transmission. This whole process is called virtual carrier sensing, which allows the area around the sender and the receiver to be reserved for communication, thus avoiding the hidden terminal problem.

### 3.2. IEEE 802.11 Power Saving Mechanism (PSM)

In this section, the IEEE 802.11 PSM is explained. The idea is to let the nodes enter a low-power doze mode if they do not receive packets. This solves the energy waste due to idle listening. In doze mode, a node consumes much less energy compared to normal mode, but cannot send or receive packets. In IEEE 802.11 PSM, this power management is done based on ad hoc traffic indication messages (ATIM). The time is divided into beacon intervals, and every node in the network is synchronized by periodic beacon transmissions. This means that each node starts and finishes each beacon interval at about the same time.

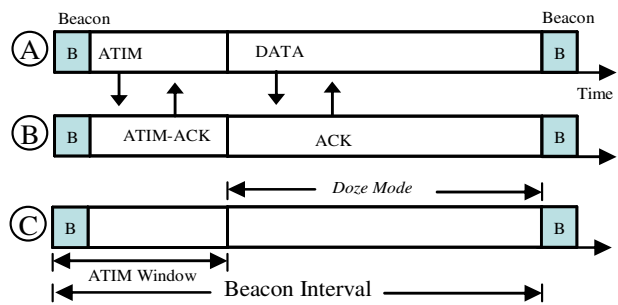

Figure 1. Operation of IEEE 802.11 PSM

Figure 1 illustrates the process of IEEE 802.11 PSM. At the start of the beacon interval, a small time frame, i.e., ATIM window is reserved for the exchange of ATIM/ATIM-ACK handshakes. Every node should be awake during this window. If node A has packets buffered for node B, then it sends an ATIM frame to B during the ATIM window. When B receives the packet, it replies with an ATIM-ACK frame. Both A and B then stay awake during the entire beacon interval. Nodes that did not send or receive an ATIM frame enter a doze mode until the next beacon interval.

### 3.3. Multichannel Hidden Terminal Problem

When a node is neither transmitting nor receiving, it listens to the control channel. When node A wants to transmit a packet to node B, A and B exchange RTS and CTS messages to reserve the channel as in IEEE 802.11 DCF. RTS and CTS messages are sent on the control channel. When sending an RTS, node A includes a list of channels it is willing to use. Upon receiving the RTS, B selects a channel and includes the selected channel in the CTS. After that, node A and B switch their channels to the agreed data channel and exchange the DATA and ACK packets.

Now consider the scenario in figure 2. Node A has a packet for B, so A sends an RTS on Channel 0 which is the control channel. B selects Channel 1 for data communication and sends CTS back to A. The RTS and CTS messages should reserve Channel 1 within the transmission ranges of A and B; so that no collision will occur. However, when node B sent the CTS to A, node C was busy receiving on another channel, so it did not hear the CTS. Not knowing that B is receiving on Channel 1, C might initiate a communication with D, and end up selecting Channel 1 for communication. This will result in collision at node B. The above problem occurs due to the fact that nodes may listen to different channels, which makes it difficult to use virtual carrier sensing to avoid the hidden terminal problem. If there was only one channel that every node listens to, C would have heard the CTS and thus deferred its transmission. Thus, we call the above problem the multichannel hidden terminal problem. As presented in the section 5, we solve this problem using synchronization, similar to IEEE 802.11 PSM.

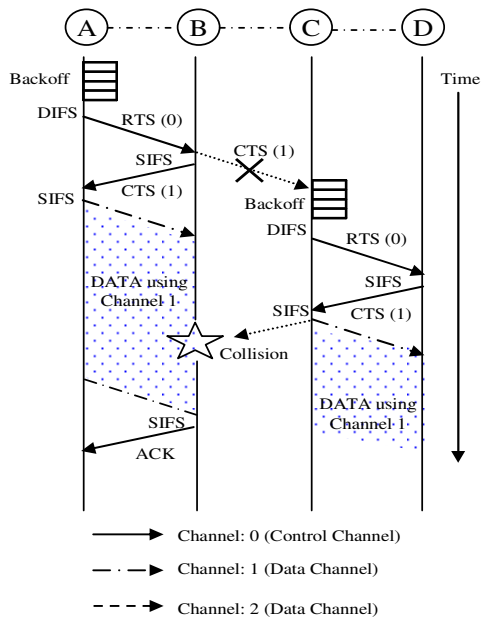

Figure 2. Multichannel hidden terminal problem

## 4. SYSTEM MODEL

We consider a multi-hop CRANs composed of a set of CR users, each of which is equipped with a single half-duplex CR transceiver. We assume CR users are stationary or moving very slowly. In our CRN, PUs are also assumed to be stationary and they coexist with the CR users. Each PU operates with an ON–OFF switching cycle that is unknown to the CRN. Consider the spectrum consisting of $C$ non-overlapping channels, each with bandwidth $B_c$ ($c = 1, 2, …, C$). These $C$ channels are licensed to PUs. CR can dynamically access any one channel to deliver its packets. Considering the fact that the spectrum opportunity is changing frequently with time and locations, we assume that CR users exchange control information in a dedicated channel which is always available. This dedicated channel may be owned by the CR service provider [28].

We assume that each transceiver always transmits at a fixed transmission power and hence, their transmission range $R_c$ and interference range $I_c$, which is typically 2 to 3 times of transmission range [29], are fixed for a particular channel $c$. We use a communication graph $G(V, E)$, to model the network where each node $v \in V$ corresponds to a CR user in the network and $E$ is the set of communication links each connecting a pair of nodes. There is a link $l = (u, v) \in E$ between nodes $u$ and $v$, if two nodes are in the transmission range and there is an available channel $c \in C_u \bigcap C_v$. Where $C_u$ and $C_v$ represent list of available channels at node $u$ and $v$ respectively. A communication link $l = (u, v)$ denotes that $u$ can transmit directly to $v$ if there are no other interfering transmissions. Due to the broadcast nature of the wireless links, transmission along a link may interfere with other link transmissions when transmitted on the same channel but links on different channels do not interfere.

An interference model defines which set of links can be active simultaneously without interfering. We model the impact of interference by using the well known protocol model proposed in [30]. A transmission on channel $c$ through link $l$ is successful if all interferes in the neighbourhood of both nodes $u$ and $v$ are silent on channel $c$ for the duration of the transmission. Two wireless links $(u, v)$ and $(x, y)$ interfere with other if they work on the same channel and any of the following is true: $v = x$, $u = y$, $v \in Nb(x)$, or $u \in Nb(y)$. Where $Nb(v)$ represents the set of neighbors of node $v$. If links $(u, v)$ and $(x, y)$ are conflicting, nodes $u$ and $y$ are within two-hops of each other [31]. The interference model can be represented by a conflict graph $F$ whose vertices corresponds to the links in the communication graph, $G$. There is an edge between two vertices in $F$ if the corresponding links can not be active simultaneously. Two links sharing a common node conflict with each other, and will have an edge in between. In addition, links in close proximity will interfere with each other if they are assigned with the same channel and hence connected with edges.

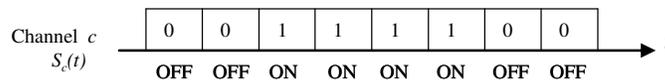

Figure 3. Channel state for the $c$-th channel

## 5. CR-MAC DESIGN

A TDMA scheme is used in the communication window of our proposed CR-MAC as depicted in the figure 4. The CR-MAC scheme has some similarities with TMMAC [32]. We assume that time domain is divided into fixed length frames and each frame consists of a sensing window, an ad hoc traffic indication messages (ATIM) window, and a communication widow. The ATIM window is contention-based and uses the same mechanism as in the IEEE 802.11 DCF [9]. The ATIM window is divided into the beacon and the control window. During the ATIM window, control channel is used for beaconing and to exchange control message. All of the CR users are synchronized by periodic beacon transmissions. In this MAC scheme, channel sensing is performed in the starting of each frame to avoid possible collisions with PUs. If any chosen channel is found to be busy, that channel will not used in the ATIM window.

As mentioned earlier, the communication window is time-slotted and uses TDMA scheme. The duration of each timeslot is the time required to transmit or receive a single data packet and it depends on the data rate of PHY layer and the size of data unit. In order to minimize possible collision with transmission from PUs, the slot size is restricted for a single data packet. The duration of the timeslot is long enough to accommodate a data packet transmission, including the time need to switch the channel, transmit the data packet and the acknowledgement. According to our MAC structure, the duration of each slot is $D_{slot} = D_{data} + D_{ACK} + 2 \times D_{guard}$. The use of guard period is to accommodate the propagation delay and the transition time from $T_x$ mode to $R_x$ mode. In the communication window, nodes can send or receive packets or go to sleep mode to save power.

If a node has negotiated to send or receive a packet in the $j^{th}$ time slot, it first switches to the negotiated channel and transmits or waits for the data packet in that slot. If a receiver receives a unicast packet, the receiver sends back an ACK in the same time slot as shown in the slot structure of figure 4. Note that proposed CR-MAC scheme does not guarantee 100% collision-free communication in the communication window, since packet collision may occur in the ATIM window which may convey incorrect information of negotiation. If a sender does not hear an ACK after it sends a unicast packet, may be because of the collision with other transmissions, the sender may perform random backoff before attempting its retransmission using free time slots. If the number of retransmissions exceeds the retry limit, the packet is dropped. It is noted that along with other channels control channel can also be used for data transmission in the communication window as shown in figure 5, if needed. If a node has not negotiated to send or receive a data packet in the $j^{th}$ time slot, the node switches to doze mode for power saving.

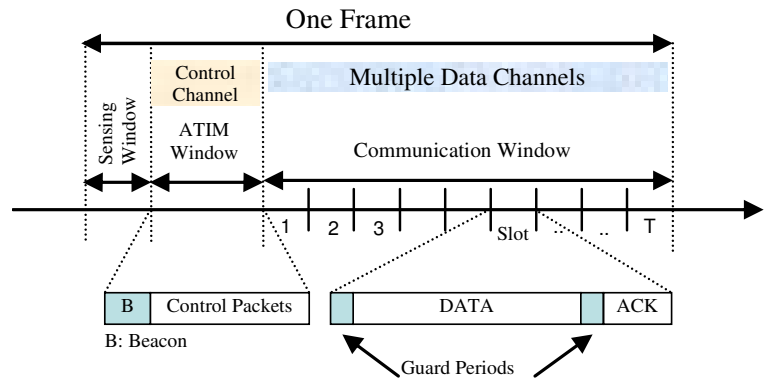

Figure 4. Structure of CR-MAC protocol

To assure collision-free communications, all neighbourhood nodes of the intended receiver except the intended transmitter should remain silent on the particular channel during a given timeslot. With the help of periodic beaconing, each node is aware of (1) the identities and list of available channels within its two-hop neighbour, and (2) existing transmission schedule of communication segments of its one-hop neighbour. Based on the collected neighbour information and its own information each secondary node update the status of its communication segments as occupied or free. Free communication segment of node *v*, *free_segment(v)*, is defined as the communication segments for all available channels, which are not used by node *v* to communicate with adjacent nodes, and are not interfere by other transmissions. Status of the communication segments on a link is determined by finding the intersection of the status of both end nodes of the link.

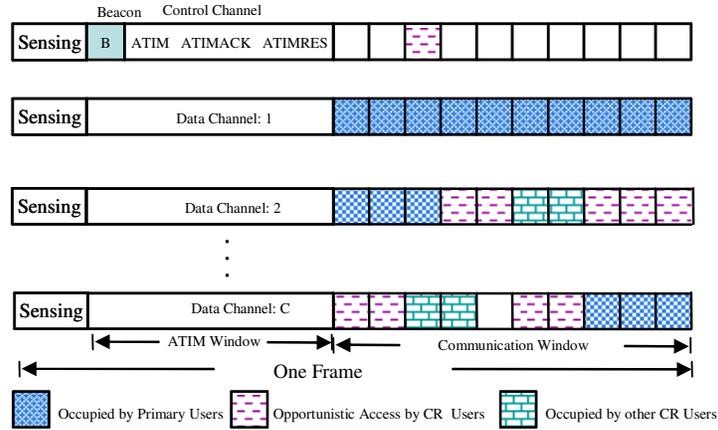

Figure 5. Process of channel negotiation and data exchange in CR-MAC

For each link in the network, the communication segment assignment algorithm marks each communication segment as one of the following:

- Occupied: this segment is using by other transmissions and hence can not be used.
- Free: unassigned idle segment.
- Assigned: this segment shall be used for packet transmission on a specific link.

We define the set of common free communication segments between two nodes to be the link bandwidth. If we let $B(u, v)$ be the available bandwidth of the link between nodes $u$ and $v$ then $B(u, v) = free\_segment(u) \bigcap free\_segment(v)$.

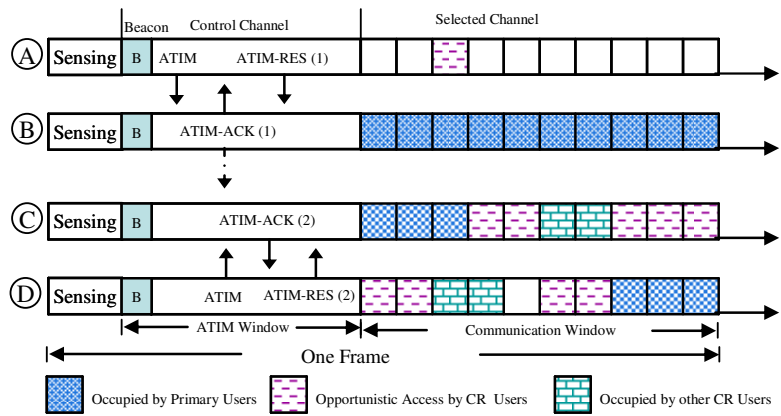

Figure 6. Solution of multichannel hidden terminal problem using CR-MAC protocol

Suppose that node A has packets for B and thus A sends an ATIM packet to B during the ATIM window, with A's free communication segment list included in the packet. On receiving the ATIM request from A, B decides which segment(s) to use during the frame interval, based on its free communication segments and A's communication segments. The communication segment (channel-timeslot) selection procedure is discussed in the next sub section. After selecting the channel and time slot(s), B sends an ATIM-ACK packet to A, specifying the channel and time slot(s) it has chosen. When A receives the ATIM-ACK packet, A will see if it can also select the channel-timeslot specified in the ATIM-ACK packet. If it can, it will send an ATIM-RES packet to B, with A's selected channel-timeslot specified in the packet. If A cannot select the channel-timeslot which B has chosen, it does not send an ATIM-RES packet to B. The process of channel-timeslot negotiation and data exchange in CR-MAC is illustrated in figure 5. Figure 6 shows how multichannel hidden terminal problem can be solved by using our CR-MAC protocol. During the ATIM window, A sends ATIM to B and B replies with ATIM-ACK indicating to use channel 1 and timeslot(s). This ATIM-ACK is overheard by C, so channel 1 will not be selected by C. When D sends ATIM to C, C selects channel 2 and timeslot(s). So, after the ATIM window, the two communications (between A and B, and C and D) can take place simultaneously in communication window.

## 5.1. Selection of Communication Segments

In this subsection, we present a heuristic algorithm to select communication segments for the link $l = (u, v)$. Let us consider $r_r(z)$ be the remaining data rate requirement for the session $z$ of a connection request. Initially $r_r(z) = r(z)$. The basic idea of this approach is to select minimum number of free communication segments to satisfy the given rate requirement within the interference constraint. In order to maintain minimum number of communication segments in a link we will use high capacity segments. Sort all the free communication segments in the descending order of their capacities. Pick a communication segment $(c, t)$ from the sorted list and check the capacity of the chosen segment $\omega(c, t) = B_c / |T|$ is not less than the $r_r(z)$, then it is selected. The selected segment is then removed from the free segment list and update the remaining rate requirement $r_r(z)$. To ensure the collision-free transmissions, the following conditions must be satisfied in selecting the communication segments. Let segment $(c, t)$ is trying to assign for the link $l = (u, v)$ such that:

- Timeslot $t$ is not assigned to any link incident (connected) on node $u$,
- Timeslot $t$ is not assigned to any outgoing link from node $v$,
- Timeslot $t$ is not used on channel $c$ by any link $l'$, $T_x(l') \in Nb(v)$, where $Nb(v)$ represents the set of neighbors of node $v$; and
- Timeslot $t$ is not used on channel $c$ by any link $l'$, $R_x(l') \in Nb(u)$.

Without confusions, $T_x(\cdot)/R_x(\cdot)$ represent both the transmitter/receiver of the given link. Note that one of the necessary constraints for collision-free communication is that no two links incident at node can be assigned same timeslot [31]. If all the above conditions are satisfied, communication segment $(c, t)$ is assigned to the link $l = (u, v)$. This procedure continues until the rate requirement is satisfied.

## 6. PERFORMANCE EVALUATION

The effectiveness of the proposed ESAMR approach is validated through simulation. This section describes the simulation environment, performance metrics, and experimental results. The result of our approach is compared with SYN-MAC [27], SCA-MAC [26], CRM-MAC [21], and IEEE 802.11 [9]. We used network simulator-2 (NS-2) version *ns-2.33* [33] to evaluate the performance of the proposed routing protocol. We generate 15 random topologies, and the result is the average over the 15 random topologies. The simulated network is composed of 50 static CR nodes deployed randomly within a $1000m \times 750m$ rectangular region. Based on the IEEE 802.11a standard, the number of channels is set to 12 including 11 data channels and one control channel. The data channels are divided into three groups that include 3 channels in the first group and 4 channels each in last two groups. Based on the IEEE 802.11b, data rates for these groups are set to 2 Mbps, 5.5 Mbps, and 11 Mbps. Nodes can respectively transmit 1, 3, or 5 consecutive packets depending on their channel condition. The data rate for control channel is 2 Mbps.

The transmission and interference range of each CR user (node) is approximately 150$m$ and 300$m$ respectively. The control channel can support a transmission range of 200$m$. We set initial energy as 60 joules per node. The number of timeslots in the communication window is set to 20 and the length of the ATIM window is 20$ms$. Channel switching delay for CR transceiver is 40$\mu s$. We randomly placed 5 PUs in the region. Each of them randomly chose a channel to use, which is then considered to be unavailable for all the CR users within their coverage range, which is set to 300$m$. We initiate sessions between randomly selected but disjoint source-destination pairs. The two-ray-ground reflection model is used to propagation model. The maximum transmission power is set to $P_{max} = 300mW$. The thermal noise power is set to $N_0 = -90dBm$. The SINR threshold is set to $\beta = 10dB$. The channel gain, $G_{uv}$ is set to $1/d_{uv}^4$, where $d_{uv}$ is the Euclidean distance between node $u$ and node $v$. The traffic demand for each communication session is given by a random number uniformly distributed in $[0.1B_c, 0.6B_c]$, where $B_c$ is the channel capacity of channel $c$. The packet size of each flow is set to 1000 bytes (excluding the sixe of IP layer and MAC layer headers). Data traffic was generated using constant bit rate (CBR) traffic sources generating 4 packets/second. All traffic sessions are established at random times near the beginning of the simulation run and they stay active until the end. Simulations are run for 500 simulated seconds. The following performance metrics are used to evaluate the proposed protocol:

*Normalized Throughput:* The ratio of throughput obtains using CRN routing protocols to the throughput obtain when using IEEE 802.11 on a single channel environment. The normalized throughput quantifies the performance improvement of CRN (multi-channel) protocols with respect to a single channel network.

*Average End-to-End Delay:* Average latency incurred by the data packets between their generation time and their arrival time at the destination.

*Packet Delivery Ratio:* The ratio of the data packets delivered to the destination to those generated by the sources.

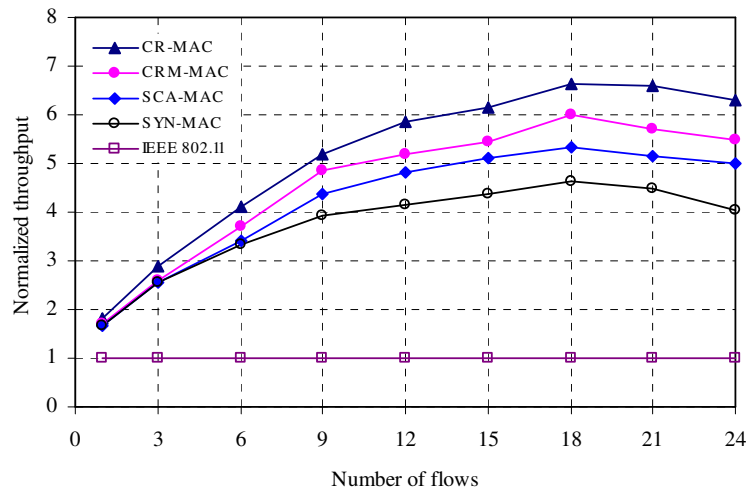

Figure 7. Network throughput varying number of flows

In the first simulation, we measured the normalized throughput varying the number of flows shown in figure 7. The throughput of CR-MAC is compared with other protocols including IEEE 802.11 single channel network using UDP traffic. The number of simultaneous UDP flows is varied from 1 to 24. As we can see from the figure, when the number of flows increases, CR-MAC offers significantly better performance than all other protocols especially compared with IEEE 802.11 single channel network. The throughput of CR-MAC is 7.4 times that of IEEE 802.11. When the network is overloaded, CR-MAC achieves 8% more throughput than CRM-MAC, 26% more than SCA-MAC, and 72% more than SYN-MAC protocol. Throughput of SYN-MAC is less because there is no CCC for conveying the control messages. As a result many connection requests are dropped resulting less throughput.

In addition, when the number of flows is large, the available channel diversity can be better exploited. Furthermore, when the number of flows is increased, CR-MAC can significantly improve the network throughput. That's because the channel assignment algorithm can balance the channel load. So the traffic is allocated on different channels in an approximate average manner. Finally, CR-MAC achieves higher performance because CR-MAC eliminates inter-flow and intra-flow interference using a non-conflicting channel-timeslot assignment.

Figure 8 shows the average end-to-end packet delay of the protocols as the network load increases. The difference between IEEE 802.11 and other protocols in delay is due to the fact that with only one channel, a packet has to wait longer to use the channel when the network load is high. When comparing with other protocols CR-MAC shows lower delay in all network scenarios. IEEE 802.11 achieves better performance than other schemes when the number of flows is less. However, according to increase of number of flows, queuing delay is raised. The queuing delay makes the performance of each protocol worse. Specially, the end-to-end packet transmission delay of IEEE 802.11 is increased dramatically according to increase of flows because IEEE 802.11 uses only a single channel for every data transmission. On the other hand, the data traffic is split into multiple channels in the case of CR-MAC. Therefore the end-to-end packet transmission delay of CR-MAC is increased slowly according to increase of flows.

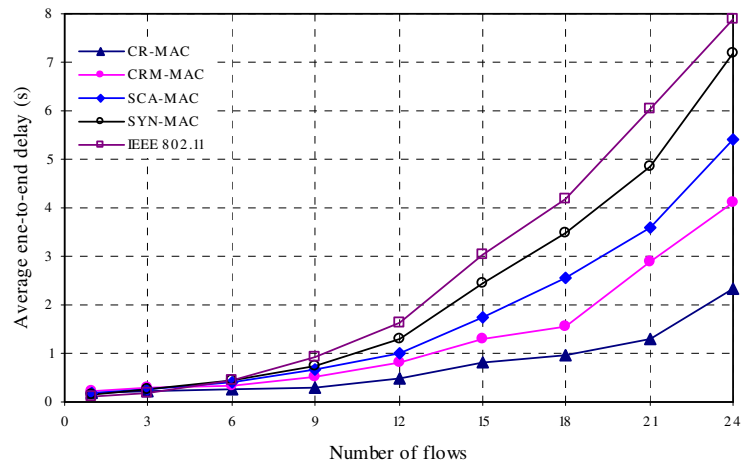

Figure 8.  Average end-to-end delay varying number of flows

Figure 9 shows the packet delivery ratio of the protocols as the network load increases. With the increasing traffic load, the burden of nodes will be aggravated and in turn resulting the failure of resource reserving. As a result, the performances of other protocols are getting worse. By using collision-free traffic scheduling, CR-MAC can balance the traffic load to different channels, which is responsible for avoiding the packets collision. Hence, CR-MAC can always keep higher packet delivery ratio.

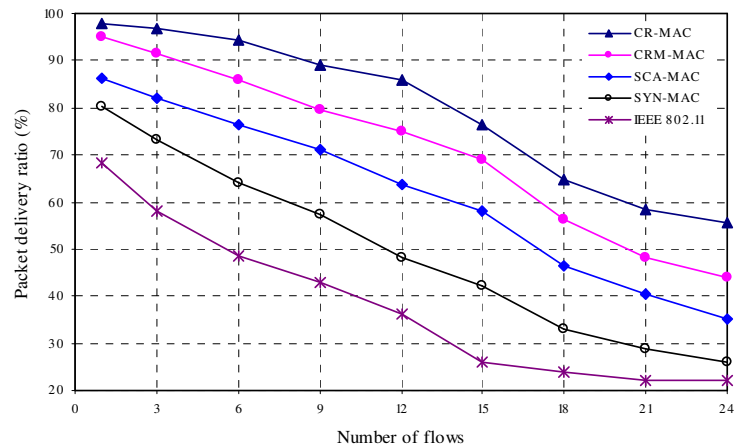

Figure 9.  Packet delivery ratio varying number of flows

## 6. CONCLUSION

In this paper, we present the CR-MAC protocol, which is a multichannel MAC protocol using a single half duplex transceiver for cognitive radio ad hoc networks. CR-MAC requires time synchronization in the network in order to avoid the multichannel hidden terminal problem and divides time into fixed frame intervals. Nodes that have packets to transmit negotiate which channels and time slots to use for data communication with their destinations during the ATIM window. This two-dimensional negotiation enables CR-MAC to exploit the advantage of both multiple channels and TDMA in an efficient way. Further, CR-MAC is able to support broadcast in an effective way.

Simulation results show that CR-MAC successfully exploits multiple channels to improve total network throughput and the end-to-end delay. Since CR-MAC only requires one transceiver per node, it can be implemented with hardware complexity comparable to IEEE 802.11. Also, power saving mechanism used in IEEE 802.11 can easily be integrated with CR-MAC for energy efficiency, without further overhead. Extensive simulations confirm the efficiency of CR-MAC and demonstrate its capability to provide high throughput for robust multi-hop communications. CR-MAC is ideal for communications under unknown and dynamic spectrum conditions, i.e. disaster recovery or military operations.

## ACKNOWLEDGEMENTS

The author would like to thank Professor Dong Geun Jeong for his valuable time and suggestions during the completion of this research work.

**Author**

**S. M. Kamruzzaman** received the B. Sc. Engineering degree in Electrical and Electronic Engineering from Dhaka University of Engineering and Technology (DUET), Bangladesh in 1997 and the M. Sc. Engineering degree in Computer Science and Engineering from Bangladesh University of Engineering and Technology (BUET), Dhaka, Bangladesh in 2005. Since September 2007, he is working towards his Ph.D. degree in Mobile Communications at Hankuk University of Foreign Studies (HUFS), Korea. From March 1998 to December 2004, he was a Lecturer and an Assistant Professor with the Department of Computer Science and Engineering, International Islamic University Chittagong (IIUC), Chittagong, Bangladesh. From January 2005 to July 2006 he was an Assistant Professor with the Department of Computer Science and Engineering, Manarat International University (MIU), Dhaka, Bangladesh. In August 2006, he moved to the Department of Information and Communication Engineering as an Assistant Professor at the University of Rajshahi (RU), Bangladesh. His research interests include radio resource management, communication protocols, cognitive radio networks, and network performance evaluation.

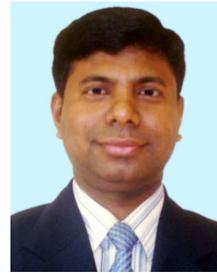